\RequirePackage[T1]{fontenc}
\documentclass[journal,10pt]{IEEEtran}
\usepackage{amsmath,amssymb,amsthm,booktabs,array,graphicx}
\usepackage{mathtools}
\usepackage[hidelinks]{hyperref}
\hypersetup{pdftitle={Frequency-Frame Covariance and Reactive-Power Scalarization},pdfauthor={Kai Sun}}
\usepackage{comment}
\newtheorem{lemma}{Lemma}
\newtheorem{theorem}{Theorem}

\newtheorem{corollary}{Corollary}
\newcommand{\ip}[2]{\langle #1,#2\rangle}
\newcommand{\norm}[1]{\lVert #1\rVert}
\newcommand{\jj}{\mathrm j}
\newcommand{\Id}{\mathrm I}
\newcommand{\Ecal}{\mathsf E}
\newcommand{\Qsig}{Q_{\Sigma}}
\newcommand{\R}{\mathbb R}

\newcommand{\PaperAuthors}{Kai Sun}
\newcommand{\PaperAffiliation}{K. Sun is with the Min H. Kao Department of Electrical Engineering and Computer Science, The University of Tennessee, Knoxville, TN 37996 USA (e-mail: kaisun@utk.edu).}

\newcommand{\ProjectionRelative}{0.272888}

\begin{document}
\title{Frequency-Frame Covariance for Reactive-Power Scalarization}
\author{\PaperAuthors\thanks{\PaperAffiliation}}
\maketitle
\begin{abstract}
Measured voltage and current waveforms may be represented in a reference frame whose frequency differs from the physical fundamental frequency. Reactive-power readouts calculated from such frequency-shifted coordinates may change if their readout operators are not transformed consistently. This paper presents a frequency-frame covariance formulation for finite-spectrum, zero-mean periodic waveforms with complete harmonic information over one physical period. A necessary and sufficient covariance condition is derived, requiring the readout operator to transform consistently with the waveform representation. Consequently, frequency-dependent harmonic weights must be evaluated at physical harmonic orders rather than relative frame labels. In particular, the derivative/Iliovici readout equals a relative-derivative contribution plus an explicit reference contribution. The formulation also separates readout-operator errors from errors caused by changes in the physical components retained by waveform projections. Three analytical cases illustrate conditional carrier recovery for a memoryless resistor, physical-frequency weighting at relative zero in an RLC branch, and the possibility that an exact reactive-power scalar can coexist with a substantial waveform error after projection. The scope is deterministic calculation consistency in frequency-shifted coordinates rather than instrument measurement uncertainty.
\end{abstract}
\begin{IEEEkeywords}
Electrical measurements, power measurement, frequency-frame covariance, complex envelopes, reactive power.
\end{IEEEkeywords}
\section{Introduction}
In AC circuits, the reactive power $Q$ of a device is calculated from its measured voltage $v$ and current $i$ according to a specified definition. For example, when only the fundamental component is considered, $Q=\operatorname{Im}{\{VI^*\}}$, where $V$ and $I$ are the corresponding RMS phasors. For nonsinusoidal waveforms, the reactive-power readout may assign different weights to the harmonic contributions, depending on the chosen definition. A measurement system may also express the voltage and current waveforms relative to a selected reference frequency. Once a reactive-power definition is selected, its reactive-power readout operator must be transformed consistently with changes in the waveform reference frame to preserve the same physical quantity.
For instance, consider a 60-Hz complex component represented in a 59.5-Hz reference. The shifted component is
\(
e^{-\jj2\pi59.5t}e^{\jj2\pi60t}=e^{\jj2\pi0.5t},
\)
with a relative or envelope frequency of 0.5 Hz, while its physical frequency remains 60 Hz. Both representations contain the same information when the reference frequency is known. However, when the waveforms are represented using a 59.5-Hz reference, the power calculation must account for that reference to preserve the physical value associated with the original waveforms. Failure to do so can introduce an error even without noise, spectral leakage, or loss of waveform coefficients.

The central question of this paper is: \emph{what must be transformed together with the frequency-shifted voltage/current representation so that a reactive-power readout continues to represent the same physical quantity?} The answer is provided by a frequency-frame covariance principle, which requires consistent transformation of waveform coordinates, the inherited physical metric, the physical-frequency mapping, the reactive-power readout operator, and any prescribed physical projection. This principle applies to measurement processing and frequency-shifted representations of AC waveforms.

IEEE Std 1459-2025 defines electrical power quantities under sinusoidal and nonsinusoidal conditions \cite{IEEE1459_2025}. This paper addresses a complementary requirement: preserving a specified reactive-power readout under an exact change of frequency reference. The analysis assumes exact voltage and current waveforms and examines reference-registration errors caused by inconsistent readout transformations. These errors are distinct from estimation error and measurement uncertainty. Methods for harmonic measurement, synchrophasor estimation, and distributed frequency estimation are discussed in \cite{Orallo2014,GarciaVeloso2024,Macii2025}. These methods, along with sensor design, analog-to-digital conversion, and hardware front-end design, are outside the scope of this paper.

Frequency-shifted representations have been used in power-system and power-electronic modeling. Generalized averaging represents a waveform through moving Fourier coefficients \cite{Sanders1991}. Dynamic-phasor models use these coefficients to describe power-system dynamics and retain the carrier terms in the transformed derivative \cite{Stankovic2000}. Carrier-shifted models and their simulation performance are studied in \cite{Dinkelbach2021}. These methods provide established representations of the waveforms and their dynamics. Here, the question is how a physical power readout should be evaluated consistently in those coordinates. This paper addresses the question for periodic waveforms with complete finite harmonic information by deriving covariance conditions for the readout operator, physical-frequency weighting, and prescribed waveform projections.

Different reactive-power definitions may serve different purposes and must remain distinct when the reference frequency frame is changed. For instance, the derivative/Iliovici scalar weights the reactive contribution of each harmonic by its order, whereas the Budeanu scalar assigns equal weights. Their different behavior for memoryless resistors is discussed in \cite{Jeltsema2014}. Related geometric-algebra power formulations organize nonsinusoidal voltage--current relationships using scalar and bivector quantities \cite{MontoyaGAPoT2020}. The related work in \cite{Sun2026} characterizes the assumptions that select a scalar reactive-power readout and quantifies the waveform information lost through that scalarization. This paper retains those distinctions and examines a different requirement: once a scalarization is selected to define the reactive-power readout, its value should remain unchanged under an exact change of representation. The same measured waveforms can legitimately give different values for different readout definitions.

To state this requirement precisely, three concepts are distinguished in this paper: \emph{covariance} transforms a specified readout together with the waveform coordinates; \emph{same-form invariance} requires the same functional to give the same value before and after a specified symmetry transformation; a \emph{device-conditioned identity} follows from a constitutive assumption, such as a known time-invariant memoryless-resistor law. The first preserves a chosen physical observable. The second can restrict the admissible functional forms. The third supplies a relation for a particular class of voltage--current pairs. Keeping these concepts distinct avoids interpreting a device-specific cancellation as a universal definition of reactive power. The paper assumes finite support on positive physical harmonics, zero-mean physical waveforms, complete and exact in-phase and quadrature (I/Q) coefficients, a common fixed real-valued carrier frequency, observation over one complete physical period, and fixed normalization. These assumptions allow the representation change to be isolated from acquisition and estimation errors and uncertainties.

Using established operator conjugation and rotating-frame derivative identities, this paper develops a frequency-frame covariance formulation for a reactive-power readout, its physical-frequency labels, and its retained waveform components. The role of operator conjugation here is to identify the physical structures that must be transported with a frequency-frame change, including the physical-frequency mapping, reactive-power readout, and any prescribed waveform projection. The main contributions are:
\begin{enumerate}
\item A necessary and sufficient condition is given for an alternating bilinear readout to retain its value for every voltage--current pair under a bijective real isometry. Applying this condition to exact frequency shifts gives physical-frequency weights and the reference correction for the derivative/Iliovici readout.

\item The covariance condition is extended to orthogonal waveform projections, such as retaining only selected harmonics. It distinguishes errors caused by an incorrectly transformed readout operator, changes in the physical components retained, and the intentional omission of waveform components. When the same physical components are retained and the readout is transformed consistently, an exact reference shift leaves both waveform and reactive-power approximation errors unchanged.
\item For a device known to be a time-invariant memoryless resistor, an identity relating its periodic voltage and current can recover an unknown reference frequency, provided that the sum of harmonic reactive-power contributions is nonzero. A sensitivity bound quantifies how data errors affect this recovery. Analytical examples involving a nonlinear resistor, an RLC branch, and waveform projections illustrate these results.
\end{enumerate}

Section II defines the waveform spaces and exact frequency frames; Section III develops readout covariance and physical-frequency weighting; Section IV treats projection covariance and error decomposition; Section V develops device-conditioned relations; Section VI presents analytical examples; and Section VII concludes the paper.

\section{Waveform Geometry and Exact Frequency Frames}
\subsection{Physical metric, orientation, and support}

This section uses exterior-algebra notation as a compact way to retain and transport multi-harmonic power information without prematurely reducing it to a scalar. Voltage and current
waveforms, $v$ and $i$, are treated as real vectors in a finite-dimensional waveform space,
while their wedge (exterior) product, $v\wedge i$, forms a bivector to organize the voltage--current information before a scalar readout is selected. On a single harmonic plane, this reduces to the familiar two-dimensional structure underlying complex phasors. Only real inner products and exterior products are needed. Throughout the paper, $(\cdot)^*$ denotes the adjoint with respect to the stated real inner product:
\(
\langle Ax,y\rangle=\langle x,A^*y\rangle.
\)
An overbar denotes complex conjugation. 

Fix $\omega_0>0$ and $T=2\pi/\omega_0$. For a finite set
$S\subset\mathbb N$ of positive physical harmonic orders, define
\begin{align}
\mathcal H_S
&=\bigoplus_{n\in S}H_n,
\qquad
H_n=\operatorname{span}_{\R}\{e_{n,c},e_{n,s}\},
\nonumber\\
e_{n,c}(t)
&=\sqrt2\cos(n\omega_0t),
\qquad
e_{n,s}(t)=\sqrt2\sin(n\omega_0t),
\label{eq:basis}\\
\ip{x}{y}
&=T^{-1}\int_0^T x(t)y(t)\,dt,
\label{eq:metric}
\end{align}
where $(e_{n,c},e_{n,s})$, denoted by $\mathcal B_n$, is the ordered, orthonormal basis of $H_n$ under \eqref{eq:metric}, and the factor $\sqrt{2}$ in \eqref{eq:basis} ensures the resulting RMS coordinates. Unless otherwise stated, all waveform inner products and reactive-power readouts in this paper are evaluated over the complete physical period $T$ in \eqref{eq:metric}. The resulting cancellation of endpoint terms is therefore part of the stated periodic waveform setting; a shorter or asynchronous observation window generally introduces additional endpoint and metric terms.

A waveform $x=\sum_{n\in S}x_n\in\mathcal H_S$ can be described by RMS coordinate columns on these planes:
\[
x_n=x_{n,c}e_{n,c}+x_{n,s}e_{n,s}
\iff
[x_n]_{\mathcal B_n}
=
\begin{bmatrix}
x_{n,c}\\
x_{n,s}
\end{bmatrix},
\]
In matrix expressions such as $v_n^TAi_m$, $v_n$ and $i_m$ denote the corresponding coordinate columns.
The common set $S$ excludes physical DC and contains all harmonics present in the voltage or current, including current-only harmonics, missing which may leave a particular scalar readout unchanged while discarding part of the current waveform information.

Define the quadrature operator $J:\mathcal H_S\rightarrow\mathcal H_S$ by the same matrix on every harmonic plane:
\begin{equation}
[J]_{\mathcal B_n}
=
\begin{pmatrix}
0&-1\\
1&0
\end{pmatrix},
\qquad n\in S .
\label{eq:J}
\end{equation}
Thus, for every $x_n\in H_n$,
\(
[Jx_n]_{\mathcal B_n}
=
[J]_{\mathcal B_n}[x_n]_{\mathcal B_n}.
\)
The basis brackets are omitted in later coordinate formulas, where $J$ denotes this $2\times2$ matrix. It rotates each coordinate column counterclockwise by $90^\circ$ and satisfies
\[
\begin{aligned}
J^*=-J
&\iff [J]_{\mathcal B_n}^{T}=-[J]_{\mathcal B_n},\\
J^2=-\Id
&\iff \bigl([J]_{\mathcal B_n}\bigr)^2=-I.
\end{aligned}
\]
For voltage and current components $v_n,i_n\in H_n$, define
\begin{equation}
Q_n
=
\ip{Jv_n}{i_n}
=
v_{n,c}i_{n,s}-v_{n,s}i_{n,c}.
\label{eq:Qn}
\end{equation}
This definition agrees with conventional harmonic reactive power. With passive current orientation, a lagging current gives positive $Q_n$. For the RMS phasors
\(
V_n=v_{n,c}-\jj v_{n,s},~
I_n=i_{n,c}-\jj i_{n,s}
\)
the conventional phasor formula gives
\[
Q_n=\operatorname{Im}(V_n\overline{I_n})
=v_{n,c}i_{n,s}-v_{n,s}i_{n,c}=\ip{Jv_n}{i_n}.
\]

To describe frequency-dependent reactive-power readouts, define the physical-order operator $N$ by \(Nx_n=nx_n\). Let
\(
E_n=e_{n,c}\wedge e_{n,s}
\)
denote the oriented unit bivector on the $n$th harmonic plane. With $D=d/dt$, differentiating the basis functions in \eqref{eq:basis} gives
\(
D=-\omega_0JN .
\)
Define:
\begin{align}
P&=\ip{v}{i},
\qquad
B=v\wedge i,
\label{eq:P_B}
\end{align}
where $P$ is the average active power and $B$ is a bivector retaining the oriented relationships between the waveform coordinates. Its norm satisfies the Gram-determinant identity
\begin{equation}
\norm B^2
=
\norm v^2\norm i^2-P^2 .
\label{eq:Bnorm}
\end{equation}
The coefficients of $B$ on the same-harmonic planes are $Q_n$. Its cross-harmonic coefficients retain other relationships already determined by the complete Fourier data \cite{MontoyaGAPoT2020,Sun2026}. 

Two established reactive-power readouts are:
\begin{align}
\Qsig
&=\ip{Jv}{i}
=\sum_{n\in S}Q_n,
\label{eq:Qs}\\
Q_D
&=-\omega_0^{-1}\ip{Dv}{i}
=\sum_{n\in S}nQ_n .
\label{eq:QDphys}
\end{align}
$Q_\Sigma$ is the equal-weight Budeanu scalar \cite{Jeltsema2014}, while $Q_D$ is the derivative/Iliovici readout \cite{Iliovici1925, Hartman2009}. The latter vanishes for all smooth, strictly monotone, time-invariant memoryless resistors under periodic excitation as proved in \cite{Sun2026}. The normalization in $Q_D$ is fixed at $\omega_0$ for the entire waveform. Consequently, a pure $n$th harmonic gives $Q_D=nQ_n$. Renormalizing separately at each harmonic would define a different readout. This fixed normalization is also retained when the processing reference changes in the rest of the paper.

\begin{table}[t]
\caption{Core notation}
\label{tab:notation}
\centering
\footnotesize
\begin{tabular}{@{}>{\raggedright\arraybackslash}p{.18\columnwidth}>{\raggedright\arraybackslash}p{.77\columnwidth}@{}}
\toprule
Symbol & Meaning\\
\midrule
$S$
& Finite set of positive physical harmonic orders.\\

$\omega_0$, $T$
& Base angular frequency and observation period,
$T=2\pi/\omega_0$.\\

$H_n$, $\mathcal H_S$
& $n$th harmonic plane and real zero-mean waveform space
$\mathcal H_S=\bigoplus_{n\in S}H_n$.\\

$\mathcal B_n$
& Ordered harmonic basis $(e_{n,c},e_{n,s})$.\\

$J$
& Counterclockwise $90^\circ$ rotation on each harmonic plane.\\

$N$
& Physical-order operator, $Nx_n=nx_n$ for $x_n\in H_n$.\\

$B=v\wedge i$
& Bivector representation of voltage--current waveform geometry.\\

$Q_n$
& Conventional reactive power of the $n$th harmonic.\\

$\Qsig$
& Equal-weight reactive-power readout,
$\Qsig=\sum_{n\in S}Q_n$.\\

$Q_D$
& Derivative/Iliovici readout,
$Q_D=\sum_{n\in S}nQ_n$.\\

$\rho$, $\nu=n-\rho$
& Common reference-frequency shift and relative-frequency label.\\

$M_\rho$
& Relative-order operator associated with the shifted frequency frame.\\

$\mathcal K_{S-\rho}$
& Support-dependent shifted-frequency image, represented as a real I/Q space.\\
\bottomrule
\end{tabular}
\end{table}

\vspace{-0.2in}
\subsection{Complex encoding and exact reference changes}
Set \(\phi_\alpha(t)=e^{\jj\alpha\omega_0t}\) for real \(\alpha\). For each harmonic plane \(H_n\), define the real-linear encoding map
\(
\Ecal:H_n\rightarrow \operatorname{span}_{\mathbb C}\{\phi_n\}
\)
by
\begin{equation}
\Ecal e_{n,c}=\phi_n,
\qquad
\Ecal e_{n,s}=-\jj\phi_n .
\label{eq:E}
\end{equation}
This map combines the cosine and sine RMS coordinates of each harmonic plane into a single complex coefficient:
\[
x_{n,c}e_{n,c}+x_{n,s}e_{n,s}
\;\xmapsto{\;\Ecal\;}\;
(x_{n,c}-\jj x_{n,s})\phi_n ,
\]
Treat the resulting complex coefficient space as a real vector space with the following inner product, which preserves the original waveform inner product:
\begin{equation}
\ip{z}{w}_{\R}
=
\operatorname{Re}\frac{1}{T}
\int_0^T \overline{z(t)}\,w(t)\,dt .
\label{eq:IQmetric}
\end{equation}
With this metric, $\Ecal$ is an isometry: \(
\ip{\Ecal x}{\Ecal y}_{\R}=\ip{x}{y}.
\)

Let $\omega_c=\rho\omega_0$ be the fixed processing reference, also called the carrier, where $\rho\in\R$ specifies the center of the frequency frame. For example, \(\rho=1\) corresponds to a frame rotating at the fundamental frequency, in which the fundamental component appears at zero relative frequency.  Define
\begin{align}
U_\rho z&=e^{-\jj\rho\omega_0t}z,\quad F_\rho=U_\rho\Ecal,\label{eq:F}\\
\mathcal K_{S-\rho}
&=\operatorname{span}_{\R}\{\phi_{n-\rho},-\jj\phi_{n-\rho}:n\in S\}.
\label{eq:image}
\end{align}
Here \(S-\rho:=\{s-\rho:s\in S\}\) denotes the frequency support \(S\) shifted by \(-\rho\). The map $F_\rho:\mathcal H_S\to\mathcal K_{S-\rho}$ therefore sends a real waveform to its shifted complex representation. For $x\in\mathcal H_S$, write $z=\Ecal x$ and $z_\rho=U_\rho z=F_\rho x$. The physical waveform is recovered by restoring the carrier before taking the real part:
\begin{equation}
x(t)=\sqrt2\operatorname{Re}\{e^{\jj\rho\omega_0t}z_\rho(t)\},
\quad F_\rho^{-1}=\Ecal^{-1}U_{-\rho}.\label{eq:inverse}
\end{equation}
The map $U_\rho$ acts on complex-valued functions, while $F_\rho$ includes the encoding of the original real waveform. Each image has dimension $|S|$ over $\mathbb C$, or $2|S|$ over $\mathbb R$. In particular, the I/Q basis directions $\phi_n$ and $-\jj\phi_n$ are linearly independent over $\mathbb R$, so their exterior product is a nonzero bivector representing the harmonic plane. In contrast, over $\mathbb C$, they are linearly dependent, so their exterior product vanishes.

A reference at the fundamental frequency has $\rho=1$. It moves the physical fundamental to zero relative frequency, but \eqref{eq:inverse} still reconstructs a physical AC component. To calculate derivatives and frequency weights in this frame, the relative order and the physical order must be distinguished. In the following lemma, $M_\rho$ returns the relative order $\nu=n-\rho$, and $N_{\mathrm{phys}}^{(\rho)}$ returns the physical order $n$. Similarly, $D_{\mathrm{env}}$ differentiates the shifted representation, whereas $D_{\mathrm{phys}}^{(\rho)}$ represents the derivative of the physical waveform.

\begin{lemma}[Exact real-carrier representation]\label{lem:frame}
$F_\rho:\mathcal H_S\to\mathcal K_{S-\rho}$ is a bijective real isometry onto its declared image. Let $\nu=n-\rho$ and define $M_\rho\phi_\nu=\nu\phi_\nu$ real linearly on that image. The transported physical operators are
\begin{align}
J_\rho&=F_\rho JF_\rho^{-1}=-\jj\Id,\nonumber\\
N_{\rm phys}^{(\rho)}&=M_\rho+\rho\Id,\label{eq:Nphys}\\
D_{\rm phys}^{(\rho)}&=F_\rho DF_\rho^{-1}
=D_{\rm env}+\jj\rho\omega_0\Id.\label{eq:connection}
\end{align}
Here $D_{\rm env}=d/dt$ on $\mathcal K_{S-\rho}$. The inner product, both RMS norms, and $\norm B$ are preserved.
\end{lemma}

\begin{proof}
Under \eqref{eq:IQmetric}, full-period Fourier orthogonality implies that
$\{\phi_n,-\jj\phi_n\}_{n\in S}$ is an orthonormal real basis. Hence,
$\Ecal:\mathcal H_S\to\mathcal K_S$ is a bijective real isometry, with
\(
\Ecal^{-1}z=\sqrt2\operatorname{Re}z.
\)
Since multiplication by the unit-modulus carrier preserves the inner product and
$U_\rho^{-1}=U_{-\rho}$, $U_\rho$ is also a bijective real isometry from
$\mathcal K_S$ onto $\mathcal K_{S-\rho}$. Therefore,
\(
F_\rho=U_\rho\Ecal
\)
is a bijective real isometry from $\mathcal H_S$ onto
$\mathcal K_{S-\rho}$. It shifts the frequency label from $n$ to
$\nu=n-\rho$ while preserving the two real I/Q coordinates and the
orthogonality between distinct frequency planes. Then, the transported physical-order and quadrature operators satisfy
\[
\begin{aligned}
N_{\rm phys}^{(\rho)}\phi_\nu
&=
F_\rho NF_\rho^{-1}\phi_\nu
=
(\nu+\rho)\phi_\nu
=(M_\rho+\rho)\phi\\
J_\rho
&=
F_\rho JF_\rho^{-1}
=
-\jj I,
\end{aligned}
\]
where the sign follows from the adopted complex-coordinate convention
$x_{n,c}-\jj x_{n,s}$.

Also, $\Ecal D\Ecal^{-1}=d/dt$ on the embedded complex functions. Thus, \(
D_{\rm phys}^{(\rho)}z
=F_\rho D F_\rho^{-1}z 
=U_\rho\frac{d}{dt}(U_{-\rho}z).
\)
Then, from \(
U_{-\rho}z=e^{\jj\rho\omega_0 t}z
\), there is immediately 
\[
D_{\rm phys}^{(\rho)}
=
D_{\rm env}+\jj\rho\omega_0 I.
\]

Finally, let $B_\rho = (\Lambda^2F_\rho)B=F_\rho v\wedge F_\rho i$. Because $F_\rho$ is a real isometry, its induced exterior map
$\Lambda^2F_\rho$ is also an isometry. Hence,
\(
\norm{B_\rho}=\norm B,
\)
consistent with the Gram-determinant identity \eqref{eq:Bnorm}.
\end{proof}

From Lemma~\ref{lem:frame}, the waveform information, RMS norms, and voltage--current geometry are preserved while the frequency label and derivative operator change because they refer to the shifted coordinates. At $\rho=1$, the identities become
\[
J_1=-\jj I,\qquad
N_{\mathrm{phys}}^{(1)}=M_1+I,\qquad
D_{\mathrm{phys}}^{(1)}=D_{\mathrm{env}}+\jj\omega_0 I.
\]
The physical fundamental $n=1$ therefore appears at zero relative
frequency, $M_1\phi_0=0$, while
\(
N_{\mathrm{phys}}^{(1)}\phi_0
=
(M_1+I)\phi_0
=
\phi_0
\)
retains the physical harmonic order. A constant complex envelope can therefore represent an oscillating physical waveform. The carrier term supplies the corresponding physical derivative. This is analogous to the familiar stationary components after a synchronous Park transformation; the present map acts on harmonic waveform coordinates rather than on three-phase instantaneous coordinates.

The derivative correction in \eqref{eq:connection} is the established carrier term in generalized averaging and dynamic-phasor representations \cite{Sanders1991,Stankovic2002}. For a general real $\rho$, the source and image spaces, $\mathcal K_S$ and $\mathcal K_{S-\rho}$, have different frequency labels. The map is an isometry between thes two declared spaces, not an automorphism of one fixed finite-support space, $\mathcal K_S$ or $\mathcal K_{S-\rho}$.

\subsection{Noninteger references and signed relative frequencies}
For physical period $T=2\pi/\omega_0$, every $z\in\mathcal K_{S-\rho}$ satisfies
\begin{equation}
z(t+T)=e^{-\jj2\pi\rho}z(t).\label{eq:twist}
\end{equation}
This is because each of its components with $n\in\mathbb Z$ satisfies
\[
\phi_{n-\rho}(t+T)
=
e^{\jj 2\pi(n-\rho)}\phi_{n-\rho}(t)
=
e^{-\jj 2\pi\rho}\phi_{n-\rho}(t).
\]
Thus, the envelope $z(t)$ is not $T$-periodic unless $\rho\in\mathbb Z$, and has the same phase change of $-2\pi\rho$ on every retained component after one physical period. This condition is called phase-twisted periodicity.

Full-period integrals can still be integrated by parts. For instance, consider two envelopes in the same frame as
{\small
\[
\begin{aligned}
z=\sum_{n\in S}Z_n\phi_{n-\rho},~
w=\sum_{k\in S}W_k\phi_{k-\rho}, \qquad
\overline z\,w=\sum_{n,k\in S}\overline{Z_n}W_k\phi_{k-n}
\end{aligned}
\]}
The common carrier cancels in the product. Since $k-n\in\mathbb Z$, $\overline z\,w$ is $T$-periodic. Integrating its derivative over $T$ gives
\begin{align}
\ip{D_{\rm env}z}{w}_{\R}
+\ip{z}{D_{\rm env}w}_{\R}
&=
\operatorname{Re}\frac{1}{T}
\int_0^T
\left(
\overline{\dot z}\,w+\overline z\,\dot w
\right)dt \nonumber\\
&=
\operatorname{Re}\frac{[\overline z\,w]_0^T}{T}
=0.
\label{eq:skew-adjoint}
\end{align}
Because
\(
D_{\rm env}\phi_{n-\rho}=\jj(n-\rho)\omega_0\,\phi_{n-\rho},
\)
differentiation also maps the finite envelope space into itself. \eqref{eq:skew-adjoint} therefore establishes skew-adjointness on that space. Both the common carrier and the full-period metric are used in this argument. A shorter physical window can leave a nonzero endpoint term.

The relative label \(\nu=n-\rho\) is a real number. At \(\nu=0\), the shifted AC plane retains the two real basis functions corresponding to \(\{1,-\jj\}\) according to \eqref{eq:E}. It should not be confused with the one-dimensional physical DC coordinate. Negative relative labels also retain the orientation inherited from their original harmonic planes. If both \(\nu\) and \(-\nu\) occur, they represent distinct planes relative to the reference frame rather than conjugate copies to be folded together.

Discarding a quadrature coordinate or filtering out negative relative labels therefore introduces an additional, generally lossy operation. Under the exact change considered here, physical time and $\omega_0$ are fixed, and the physical frequency remains $\omega_0(\nu+\rho)=n\omega_0$.

Successive exact reference changes are also consistent. For fixed real $\rho,\sigma,\eta$,
\begin{align}
T_{\rho\to\sigma}&=F_\sigma F_\rho^{-1}
=U_{\sigma-\rho}|_{\mathcal K_{S-\rho}},\nonumber\\
T_{\sigma\to\eta}T_{\rho\to\sigma}&=T_{\rho\to\eta}.
\label{eq:composition}
\end{align}
The second identity follows by canceling $F_\sigma^{-1}F_\sigma$. Hence, $T_{\rho\to\rho}=\Id$ and $T_{\rho\to\sigma}^{-1}=T_{\sigma\to\rho}$. Transforming data, operators, or projections directly to a final frame gives the same result as passing through an intermediate frame.

\section{Readout Covariance and Physical Weights}
\subsection{Separating waveform data from the selected readout}
A scalar reactive-power definition specifies how voltage and current waveform data are utilized, while the waveform data alone do not select that definition. 
To describe this distinction, let $H$ be a $m$-dimensional real inner-product space and let $q_A(v,i)=\ip{Av}{i}$ denote a general alternating bilinear scalar readout defined by a skew-adjoint operator $A$ satisfying $A^*=-A$. Interchanging $v$ and $i$ reverses the sign of $q_A$. Conversely, every alternating bilinear scalar readout $q$ on $H$ has such a unique representation of this form. For each $v$, the real Riesz representation theorem gives a unique vector $Av$ such that $q(v,i)=\ip{Av}{i}$ for all $i\in H$. Bilinearity makes $A$ linear, while the alternating property $q(v,i)=-q(i,v)$ implies $A^*=-A$.

The same readout can also be expressed as a contraction of two real bivectors, which separates the waveform data from the scalar readout. The inner product on $H$ induces an inner product on $\Lambda^2H$: 
\begin{equation}
\ip{x\wedge y}{u\wedge w}_{\Lambda^2}
=
\ip{x}{u}\ip{y}{w}
-
\ip{x}{w}\ip{y}{u}.
\label{eq:wedgeMetric}
\end{equation}
For a real orthonormal basis $\{e_1,\ldots,e_m\}$ of $H$, define
\begin{equation}
\kappa_A
=
\sum_{a<b}
\ip{Ae_a}{e_b}\,e_a\wedge e_b.
\label{eq:kappa}
\end{equation}
$\kappa_A$ tells which oriented waveform planes are included in the alternating scalar readout $q_A$ and defines its equivalent bivector representation; namely
\begin{equation}
q_A(v,i)
=
\ip{v\wedge i}{\kappa_A}_{\Lambda^2}.
\label{eq:contraction}
\end{equation}
Thus, bivector $B=v\wedge i$ contains the voltage--current waveform data, while $\kappa_A$ defines the scalar reactive-power readout $q_A$ applied to the data. Different choices of $\kappa_A$ can produce different scalar values from the same data and bivector.

\vspace{-0.2in}
\subsection{Condition for preserving the reported scalar}
When the representations of voltage and current are changed, the reactive-power readout must be changed accordingly. The following theorem determines how the readout operator must transform so that the same scalar value is obtained for every voltage--current input pair.

\begin{theorem}[Operator covariance under exact representation changes]\label{thm:cov}
Let $U:H\to H'$ be a bijective real isometry between finite inner-product spaces, and let $A^*=-A$. For a linear $A'$ on $H'$,
\begin{equation}
\begin{aligned}
&q_{A'}(Uv,Ui)=q_A(v,i)\quad\text{for every }v,i\\
&\hspace{2em}\Longleftrightarrow\quad A'=UAU^{-1}.
\end{aligned}\label{eq:cov}
\end{equation}
For this operator, $A'^*=-A'$, and with $U_2=\Lambda^2U$,
\begin{equation}
B'=U_2B,\qquad\kappa_{A'}=U_2\kappa_A.\label{eq:joint}
\end{equation}
\end{theorem}
\begin{proof}
Because $U$ is a bijective isometry, $U^{-1}=U^*$. Thus, if $A'=UAU^{-1}$, then
\(
(A')^* = UA^*U^{-1} = -UAU^{-1} = -A',
\)
so the transformed operator remains skew-adjoint.

To prove sufficiency, from $A'=UAU^{-1}$, $\ip{A'Uv}{Ui} = \ip{UAv}{Ui} = \ip{Av}{i},$
where the last equality follows from the isometry of $U$. Hence, the transformed readout gives the same scalar value for every voltage--current pair.

For necessity, suppose $\ip{A'Uv}{Ui} = \ip{Av}{i}$ for all $v,i$. Using the isometry of $U$, this is equivalent to $\ip{(U^{-1}A'U-A)v}{i}=0$ for all $v,i$.
For each fixed $v$, the only vector having zero inner product with every $i$ is the zero vector. Thus, $(U^{-1}A'U-A)v=0$ for all $v$, which gives $A'=UAU^{-1}$.

Now consider the exterior-product form. Let $U_2=\Lambda^2U$. The bivector on transformed data is $B'= Uv\wedge Ui = U_2B.$ 
Because $U$ is an isometry on $H$, the induced inner product in \eqref{eq:wedgeMetric} makes $U_2$ an isometry on $\Lambda^2H$.
Then, from \eqref{eq:contraction}, for every decomposable bivector $B=v\wedge i$, 
\(
\ip{U_2B}{\kappa_{A'}} = q_{A'}(Uv,Ui)
= q_A(v,i) = \ip{B}{\kappa_A}_{\Lambda^2}
= \ip{U_2B}{U_2\kappa_A}_{\Lambda^2}.
\)
Since decomposable bivectors span $\Lambda^2H$, this equality for all $B$ implies
\(
\kappa_{A'}=U_2\kappa_A,
\)
which proves the second identity in \eqref{eq:joint}.
\end{proof}
Theorem~\ref{thm:cov} gives the condition for a reactive-power readout to remain unchanged under a change of waveform representation. For \(U=F_\rho\), the readout operator must be transformed by conjugation to preserve the same scalar for every voltage--current pair. In the bivector form, \(B\) and \(\kappa_A\) transform together while their inner product remains invariant. This is an operator-level result: an incorrectly transformed operator may agree for one waveform pair if its mismatch vanishes on that input, but not in general.

If an implementation uses an operator different from the correctly transported operator $A'=UAU^{-1}$, its readout error can be expressed in the original coordinates.

\begin{corollary}[Exact operator mismatch]
If $A''$ is used in place of the correctly transported operator $A'=UAU^{-1}$, then
\begin{equation}
q_{A''}-q_A=\ip{\Delta A v}{i},
\quad
\Delta A=U^{-1}A''U-A,
\label{eq:operatorMismatch}
\end{equation}
\vspace{-0.2in}
\begin{equation}
\left|q_{A''}(Uv,Ui)-q_A(v,i)\right|
\le
\norm{\Delta A}_{\mathrm{op}}\norm{v}\norm{i}.
\label{eq:operatorMismatchBound}
\end{equation}
\end{corollary}
\noindent where $\norm{\Delta A}_{\mathrm{op}}=\sigma_{\mathrm{max}}(\Delta A)$ is the induced operator norm equal to the largest singular value of $\Delta A$, measuring the maximum mismatch over all waveform directions.

\begin{proof}
Using the isometry of $U$, the implemented readout can be expressed in the original coordinates as
\(
q_{A''}(Uv,Ui)=\ip{U^{-1}A''Uv}{i}.
\)
Subtracting $q_A(v,i)=\ip{Av}{i}$ gives \eqref{eq:operatorMismatch}. Applying the Cauchy--Schwarz inequality and the definition of the operator norm gives \eqref{eq:operatorMismatchBound}.
\end{proof}

In the rest of the paper, a double prime denotes an incorrectly transported operator, e.g. \(A''\), to distinguish it from the correctly transported denoted by a single prime, e.g. \(A'\).

A useful case is $A=JW$, where $W^*=W$ supplies the weights. The conditions $J^*=-J$ and $JW=WJ$ imply $A^*=-A$. Transforming both factors then gives $A'=J'W'$. The commutation condition matters: without it, the skew part of $JW$ is $(JW+WJ)/2$, and $JW$ does not necessarily define an alternating, scalar reactive-power readout. 

\vspace{-0.2in}
\subsection{Weights evaluated at physical frequency}
For a real weighting function $f:S\to\R$ on the physical harmonic orders, define the frequency-weighted reactive-power readout operator and its scalar readout as
\begin{equation}
A_f=Jf(N),\qquad
q_f=\ip{A_fv}{i}
=\sum_{n\in S}f(n)Q_n.
\label{eq:qf}
\end{equation}
Since $J$ and $N$ commute, $A_f$ is skew-adjoint. The following theorem gives the representation of the same readout in an arbitrary fixed frequency frame.

\begin{theorem}[Physical-frequency weights]\label{thm:weights}
Under the observation assumptions of Section II, the same scalar in every fixed real frame is
\begin{align}
&q_f=\ip{J_\rho f(M_\rho+\rho\Id)v_\rho}{i_\rho}_{\R}\nonumber
=\sum_{\nu\in S-\rho}f(\nu+\rho)Q_\nu^{(\rho)},\\
&Q_\nu^{(\rho)}=Q_{\nu+\rho}.\label{eq:weights}
\end{align}
\end{theorem}
\begin{proof}
Lemma~\ref{lem:frame} and finite spectral calculus give
$F_\rho f(N)F_\rho^{-1}=f(M_\rho+\rho\Id)$; the latter has eigenvalues only at $\nu+\rho\in S$. Transport $J$ by the same map and apply Theorem~\ref{thm:cov}. The oriented two-coordinate planes are unchanged, giving the displayed sum.
\end{proof}

Theorem~\ref{thm:weights} shows that a reference change only relabels the physical harmonic order $n$ as the relative order $\nu=n-\rho$. Therefore, the physical weight $f(n)$ must be evaluated at $n=\nu+\rho$ in the shifted frame. 
Equation~\eqref{eq:weights} is the implementation rule: preserve the physical-frequency label associated with each I/Q pair, and use that label in the selected weighting function. It also covers negative relative frequencies and relative zero, because the corresponding physical order remains positive.

Reusing the same function on relative labels is a different requirement from covariance. If \(h(\nu)\) is used unchanged after the shift \(\nu=n-\rho\), the resulting mismatch is \(\sum_{n\in S}[h(n-\rho)-h(n)]Q_n\). This mismatch vanishes for every waveform and every real \(\rho\) only when \(h\) is constant; nonconstant physical weights remain fully covariant when their argument is transformed according to Theorem~\ref{thm:weights}. 

This conclusion concerns reuse of the same relative-label rule. It does not restrict the physical weights allowed by covariance: any function \(f\) on \(S\) is admissible when its argument is transformed according to \eqref{eq:weights}. A broader classification of readouts that remain unchanged under frequency-label shifts is a stronger problem and is not pursued here.
\vspace{-0.2in}
\subsection{Correction for the derivative readout}
The derivative/Iliovici readout provides a direct example. Its physical weighting is $f(n)=n$. After the reference change $n=\nu+\rho$, covariance requires the weight $\nu+\rho$, whereas reusing only the relative weight $\nu$ omits the carrier contribution. Substituting $n=\nu+\rho$ in \eqref{eq:weights} gives
\begin{align}
Q_D&=Q_{D,\mathrm{rel}}^{(\rho)}+\rho\Qsig,
\label{eq:correction}\\
Q_{D,\mathrm{rel}}^{(\rho)}
&=-\omega_0^{-1}\ip{D_{\rm env}v_\rho}{i_\rho}_{\R}
=\sum_{\nu\in S-\rho}\nu Q_{\nu+\rho}.
\nonumber
\end{align}
The first term is calculated from the envelope derivative. The second restores the part of the physical derivative associated with the selected carrier. Both terms use the same fixed normalization $\omega_0$. If a measurement processor intends to report $Q_D$ but instead reports only $Q''_D=Q_{D,\mathrm{rel}}^{(\rho)}$, its deterministic error is
\begin{equation}
{Q''_D-Q_D=-\rho\Qsig.}\label{eq:registration-bias}
\end{equation}
The sign in \eqref{eq:registration-bias} follows from the passive orientation and normalization defined in Section II. The error is caused by using the wrong derivative operator; the physical waveform has not changed. The same result follows by substituting \eqref{eq:connection} into the derivative readout.

Physical-frequency weighting is also needed for reciprocal weights. For example, if \(f(n)=1/(2\omega_0 n)\), the denominator in the shifted frame must use the physical order \(\nu+\rho=n>0\), not \(\nu\). A harmonic that appears at \(\nu=0\) after a frame shift still has the nonzero physical order \(n=\rho\) and does not represent zero physical frequency. Substituting \(1/\nu\) would introduce a singularity.

\section{Waveform Projection and Readout Error Analysis}
Building on the exact frame and reactive-power readout covariance established above, this section addresses lossy waveform projection, which may discard physical information and make reactive-power scalarization approximate, and quantifies the resulting readout errors.

\vspace{-0.2in}
\subsection{Preserving the selected physical subspace}
A measurement processor may retain only part of the waveform data. This selection must be specified separately from the frequency reference. An exact reference frame change is lossless under the assumptions of Section II, whereas a projection can discard information.

Let \(\Pi\) be an orthogonal projection satisfying \(\Pi^2=\Pi\) and \(\Pi^*=\Pi\), which selects a physically defined set of harmonic planes or a more general subspace under the physical inner product. The reduced observation is the pair \((\Pi v,\Pi i)\) together with its specified reactive-power readout. To represent the same reduced observation across reference frames, the projection must transform with the waveform coordinates.

\begin{theorem}[Compatible projection and readout consistency]\label{thm:projection}
For a bijective real isometry $U$, a target projection represents the same physical projection if and only if
\begin{equation}
\Pi'U=U\Pi\quad\Longleftrightarrow\quad\Pi'=U\Pi U^{-1}.\label{eq:projection}
\end{equation}
With $A'=UAU^{-1}$,
\begin{equation}
q_{A'}(\Pi'Uv,\Pi'Ui)=q_A(\Pi v,\Pi i).\label{eq:projectedCov}
\end{equation}
The error relative to the full readout obeys
\begin{align}
&|q_A(v,i)-q_A(\Pi v,\Pi i)|\nonumber\\
&\quad\leq\norm A_{\rm op}\big(\norm{v-\Pi v}\norm i
+\norm{\Pi v}\norm{i-\Pi i}\big).\label{eq:projectedBound}
\end{align}
\end{theorem}
\begin{proof}
Multiply $\Pi'U=U\Pi$ on the right by $U^{-1}$; the converse is immediate. Conjugation preserves self-adjointness and idempotence, hence orthogonality. Substitute $\Pi'U=U\Pi$ into Theorem~\ref{thm:cov} to obtain \eqref{eq:projectedCov}. Add and subtract $\ip{A\Pi v}{i}$ to express the full-minus-projected difference as
\(
\ip{A(v-\Pi v)}{i}+\ip{A\Pi v}{i-\Pi i}.
\)
Cauchy--Schwarz and the operator norm give \eqref{eq:projectedBound}.
\end{proof}

Two results follow: first, properly transforming the projection preserves which physical data are retained; second, the error bound compares the resulting projected readout with the full readout and quantifies the error caused by discarding data.

With the waveform projection now defined, when the frequency frame is changed, the waveform coordinates, reactive-power readout operator, and waveform projection must all be transformed consistently so that the same physical waveform components are retained and the same reactive-power scalar is obtained. In the $\rho$-frame, set
\begin{align}
v_\rho&=F_\rho v,\quad i_\rho=F_\rho i,\nonumber\\
A_\rho&=F_\rho AF_\rho^{-1},\quad
\Pi_\rho=F_\rho\Pi F_\rho^{-1}.\label{eq:synthesis}
\end{align}
With the inherited inner products and the corresponding inner products induced on the exterior-product spaces as in \eqref{eq:wedgeMetric}, these identities follow from Lemma~\ref{lem:frame} and Theorems~\ref{thm:cov} and \ref{thm:projection}:
\begin{align}
&B_\rho=(\Lambda^2F_\rho)B,\quad
\kappa_{A_\rho}=(\Lambda^2F_\rho)\kappa_A,\nonumber\\
&q_{A_\rho}(v_\rho,i_\rho)=q_A(v,i),\nonumber \\
&q_{A_\rho}(\Pi_\rho v_\rho,\Pi_\rho i_\rho)=q_A(\Pi v,\Pi i).
\label{eq:joint-observation}
\end{align}

From~\eqref{eq:composition}, the final result depends only on the initial and final frames, not on intermediate transformations. The readout-consistency condition in \eqref{eq:cov} and Theorem~\ref{thm:cov} determines how the reactive-power operator must transform for all voltage--current pairs, whereas \eqref{eq:projection} and Theorem~\ref{thm:projection} determine how the projection must transform to preserve the same physical waveform subspace. Since different projections may retain different waveform components while producing the same scalar value, equality of the projected readout in \eqref{eq:projectedCov} for one pair may happen. Thus, such equality alone does not identify the projection or the retained physical data.

A fixed selection of relative-frequency components may correspond to different physical harmonics as $\rho$ changes, even when the retained components use the correct weight $\nu+\rho$. Transforming the projection with the frame preserves the same physical waveform components. For the derivative readout $A=JN$, \eqref{eq:projectedBound} has
\(
\norm{A}_{\rm op}=\max_{n\in S} n,
\)
because $N$ weights the $n$th harmonic by $n$ and $J$ preserves norm. This bound can be conservative because it measures waveform approximation error rather than the actual readout contribution of omitted harmonics. In particular, omitting a current-only harmonic changes the current waveform but contributes no corresponding $Q_n$, and the projected voltage--current pair does not need to satisfy the original pointwise nonlinear device law. Example E3 in Section VI illustrates both effects.

\begin{corollary}[Frame-invariant approximation errors]\label{cor:orbit}
For a fixed physical orthogonal projection $\Pi$, joint transformation of the waveforms, readout operator, and projection preserves both waveform approximation and scalar readout errors:
{\small
\begin{align}
\norm{x_\rho-\Pi_\rho x_\rho}_{\R}&=\norm{x-\Pi x},\label{eq:invariant-wave-error}\\
q_{A_\rho}(v_\rho,i_\rho)
-q_{A_\rho}(\Pi_\rho v_\rho,\Pi_\rho i_\rho)&=q_A(v,i)-q_A(\Pi v,\Pi i).\label{eq:invariant-scalar-error}
\end{align}
}
\end{corollary}
\begin{proof}
Since $x_\rho-\Pi_\rho x_\rho=F_\rho(x-\Pi x)$, so isometry proves \eqref{eq:invariant-wave-error}. Subtract the last two identities in \eqref{eq:joint-observation} for \eqref{eq:invariant-scalar-error}.
\end{proof}

Corollary~\ref{cor:orbit} shows that an exact reference change preserves the error of a fixed physical approximation. However, a readout difference between two reference frames can arise from either an error in transforming the readout operator or a change in the physical harmonics retained by the projection.

\subsection{Separating three deterministic error mechanisms}
The distinction can be made explicit when both the readout and projection are implemented incorrectly. Let $F:H\to H'$ be the declared real isometry. Let $A''$ and $\Pi''$ denote, respectively, the incorrectly implemented readout operator and projection in the transformed coordinates, while $A'$ and $\Pi'$ denote their correct transformations. Pulling the implemented projection back to the physical coordinates gives
\begin{equation}
C=F^{-1}\Pi''F.
\label{eq}
\end{equation}
Since $C^*=C=C^2$, the implemented scalar can be written in the physical coordinates as
\(
q''=\ip{C(F^{-1}A''F)Cv}{i}.
\)
Adding and subtracting the corresponding intended operations gives
\begin{equation}
\begin{aligned}
&q''-q_A(v,i)
=\big\langle (\epsilon_1+\epsilon_2+\epsilon_3)v,i\big\rangle,\\
&\epsilon_1=C(F^{-1}A''F-A)C,\\
&\epsilon_2=CAC-\Pi A\Pi, \qquad \epsilon_3=\Pi A\Pi-A.
\end{aligned}
\label{eq:error-decomposition}
\end{equation}
Here, $\epsilon_1$ represents the readout-operator mismatch on the implemented subspace, $\epsilon_2$ represents the mismatch between the implemented and intended physical projections, such as that caused by selecting the wrong physical components, and $\epsilon_3$ represents the approximation or truncation error introduced by the intended projection itself, such as that caused by intentionally discarding physical components. Correctly transforming the readout operator avoids $\epsilon_1$, but does not generally eliminate $\epsilon_2$ or $\epsilon_3$. The three errors in ~\eqref{eq:error-decomposition} may cancel and are not assumed to be statistically independent. In fact, $\epsilon_1+\epsilon_2+\epsilon_3=C(F^{-1}A''F)C-A$. Therefore, a small total scalar residual does not separately identify the three error mechanisms. This distinction is important when comparing reactive-power readouts from different measurement processors using different reference frames or harmonic selections.

\vspace{-0.1in}
\subsection{Suggested measurement-processing procedure}
For finite-dimensional periodic waveforms, the preceding results suggest this measurement-processing procedure.
\begin{enumerate}
\item Specify the physical reactive-power readout, its normalization, the observation period, and the time origin. Retain the common carrier and both I/Q coefficients for every physical harmonic in the declared support.

\item Transform the waveform coordinates and the readout together. For harmonic weighting, use the physical order $n=\nu+\rho$, not the relative order $\nu$. For example, for $Q_D$, include the carrier term in \eqref{eq:correction} to eliminate  $\epsilon_1$.

\item If the waveform data are reduced after applying a filter, transform the intended physical projection according to \eqref{eq:projection}. Separate the projection mismatch $\epsilon_2$ from the readout mismatch $\epsilon_1$ and the approximation error $\epsilon_3$.

\item Reconstruct the real waveforms before applying a physical device law, and report the selected scalarization, physical harmonic support, reference information, and any approximation used.
\end{enumerate}
This procedure addresses deterministic consistency once waveform data satisfying the stated assumptions are available. Acquisition errors and uncertainty propagation require additional models of the measurement channel and processing algorithm.

\vspace{-0.2in}
\section{Device-Conditioned Identities for Carrier Recovery and Energy Interpretation}
This section shows how device-specific physical relations can be combined with the proposed covariant readout framework to recover the carrier frequency and to interpret selected reactive-power readouts in terms of stored energy.

\vspace{-0.2in}
\subsection{Transport of a memoryless device law}
Covariance applies to a specified reactive-power readout without requiring a device model. A known device law supplies additional relations that can be used to check a calculation or recover missing reference information. Those relations must be applied to the same physical waveforms.

Consider a real memoryless law $i=g(v)$. To keep the exact finite-coordinate representation, restrict its input to
\(
\mathcal D_g=\{v\in\mathcal H_S:g(v)\in\mathcal H_S\};
\)
i.e., the common support $S$ must include all harmonics generated in the actual output. On the transformed admissible set $F_\rho\mathcal D_g$, the same device law is written as
\begin{equation}
i_\rho=(F_\rho\circ g\circ F_\rho^{-1})(v_\rho).\label{eq:device}
\end{equation}
The composition first reconstructs the real voltage, then evaluates the physical law, and finally transforms the resulting current. Evaluating $g$ directly at complex-envelope samples is not generally the same operation. Moreover, a smooth nonlinear law can generate DC or infinitely many harmonics, so it does not necessarily preserve the entire finite space $\mathcal H_S$. The time-domain identity below has a broader domain; its finite-coordinate form requires the stated admissibility conditions.
\vspace{-0.2in}
\subsection{Conditional recovery of missing carrier information}
For a time-invariant memoryless resistor, the physical derivative reactive-power readout vanishes over a complete period \cite{Sun2026}. Combining this relation with the carrier correction gives a scalar constraint on the reference. The constraint identifies the carrier only when the excitation makes its coefficient nonzero.

\begin{theorem}[Periodic resistor diagnostic and conditional carrier recovery]\label{thm:resistor}
Let $v$ be real, $C^1$, and $T$-periodic, and let $i=g(v)$ for continuous time-invariant $g$. Then
\begin{equation}
-\frac1{\omega_0T}\int_0^T\dot v\,g(v)\,dt=0.\label{eq:chain}
\end{equation}
For pairs also satisfying the finite I/Q assumptions, every fixed real carrier satisfies
\begin{equation}
Q_{D,\mathrm{rel}}^{(\rho)}+\rho\Qsig=0.\label{eq:resistor}
\end{equation}
Under this device prior, complete data, fixed $\omega_0$, and a common unknown real carrier, $\Qsig\ne0$ permits recovery of its missing reference information:
\begin{equation}
\rho=-\frac{Q_{D,\mathrm{rel}}^{(\rho)}}{\Qsig}.\label{eq:recovery}
\end{equation}
If $\Qsig=0$, this scalar constraint supplies no carrier information. For perturbations $\delta q$, $\delta s$ in the relative derivative readout and $\Qsig$, with $|\delta s|<|\Qsig|$,
\begin{equation}
|\widehat\rho-\rho|\leq
\frac{|\delta q|+|\rho|\,|\delta s|}{|\Qsig|-|\delta s|}.\label{eq:sensitivity}
\end{equation}
\end{theorem}

\begin{proof}
Let $G'=g$. Periodicity gives
\[
\int_0^T \dot v\,g(v)\,dt = \int_0^T \frac{d}{dt}G(v)dt
=G(v(T))-G(v(0))=0.
\]
Thus, the physical derivative reactive-power readout vanishes. For finite admissible pairs, substituting this result into \eqref{eq:correction} gives \eqref{eq:resistor}. If $\Qsig\ne0$, solving for $\rho$ gives \eqref{eq:recovery}; if $\Qsig=0$, \eqref{eq:resistor} provides no information about $\rho$. For the sensitivity bound, let $s=\Qsig$ and $q=-\rho s$. Then
\[
\widehat\rho-\rho
=-\frac{q+\delta q}{s+\delta s}-\rho
=\frac{-\delta q-\rho\delta s}{s+\delta s}.
\]
Hence,
\[
|\widehat\rho-\rho|
\le
\frac{|\delta q|+|\rho||\delta s|}
{|s|-|\delta s|},
\]
where $|s+\delta s|\ge |s|-|\delta s|$ and $|\delta s|<|s|$. This proves \eqref{eq:sensitivity}.
\end{proof}

The function $G$ is only an antiderivative of the constitutive law and is not a stored-energy function. The zero derivative readout follows from periodicity and does not require monotonicity; monotonicity is relevant only to passive device classes and the converse result in \cite{Sun2026}.
Equation~\eqref{eq:recovery} combines this device identity with the carrier correction to recover $\rho$. This requires $\Qsig\ne0$: when $\Qsig=0$, the scalar relation provides no carrier information, while a small nonzero $\Qsig$ makes the recovery sensitive to perturbations as quantified by \eqref{eq:sensitivity}.
Therefore, the recovery requires a known periodic time-invariant memoryless-resistor law, complete admissible data, fixed physical normalization, one common carrier, and $\Qsig\ne0$. It does not identify an arbitrary unknown load or prove that a device is memoryless. The recovered carrier is real and should not be rounded without an integer-carrier prior. 
\vspace{-0.2in}
\subsection{Signed mean-storage imbalance in a periodic RLC branch}
A second device-specific interpretation is available for a linear time-invariant (LTI) series RLC branch in periodic steady state. Let $I_n$ be the RMS current phasor at physical order $n$. The voltage and mean stored energies are
\begin{align}
V_n&=\left[R+\jj\left(n\omega_0L-\frac1{n\omega_0C}\right)\right]I_n,\nonumber\\
W_{L,n}&=\tfrac12L|I_n|^2,\quad
W_{C,n}=\frac{|I_n|^2}{2C(n\omega_0)^2}.\label{eq:energy}
\end{align}
Here $W_{L,n}$ and $W_{C,n}$ are the full-period mean magnetic and electric energies at the $n$th harmonic. Substitution into the phasor power formula gives its reactive-power contribution:
\begin{equation}
Q_n=2n\omega_0(W_{L,n}-W_{C,n}).\label{eq:energyRelation}
\end{equation}
which depends on the signed difference between the two mean energies \cite{Sun2026}. At series resonance, the difference can be zero while both mean energies remain positive.

To retain the per-harmonic signed differences, define a linear map on bivectors by $\mathcal R_f E_n=f(n)E_n$ that multiplies the coefficient of $E_n=e_{n,c}\wedge e_{n,s}$ by physical weight $f(n)$ and the cross-harmonic basis elements by zero. Thus, only components associated with $Q_n$ are retained and weighted by $f(n)$. Let \(f(n)=\frac{1}{2\omega_0 n}\). The storage-imbalance bivector is then
\begin{equation}
\mathcal E_\Delta=\mathcal R_f B
=\sum_n\frac{Q_n}{2\omega_0n}E_n=\sum_n (W_{L,n}-W_{C,n})E_n.\label{eq:energyBivector}
\end{equation}

After a frequency-frame change defined by $F_\rho$, there is
\begin{equation}
\mathcal R_f^{(\rho)}=(\Lambda^2F_\rho) \mathcal R_f (\Lambda^2F_\rho)^{-1}.
\label{eq:energyTransport}
\end{equation}
The resulting energy-imbalance coefficient must be $\frac{Q_\nu^{(\rho)}}{2\omega_0(\nu+\rho)}$ using physical frequency, not $\frac{Q_\nu^{(\rho)}}{2\omega_0\nu}$. This energy interpretation is specific to the stated periodic LTI branch model and does not establish a general nonlinear storage relation.

\vspace{-0.1in}
\section{Illustrative Cases}
This section uses three analytical cases to illustrate the covariance formulation. The first examines reference correction for a nonlinear resistor, the second shows physical-frequency weighting at relative zero in an RLC branch, and the third distinguishes correct frequency weighting from correct waveform selection. The cases also illustrate the different error mechanisms identified in \eqref{eq:error-decomposition}.

All amplitudes and component values are dimensionless normalized quantities. The reference values are obtained directly from the displayed analytic waveforms and are independently checked by numerical sampling and RMS-coefficient extraction. These checks verify the formulas and their implementation. The cases include both integer reference shifts and the fractional shift $\rho=119/120$, corresponding to the 60-Hz/59.5-Hz example in the Introduction.

\vspace{-0.2in}
\subsection{E1: nonlinear-resistor readouts and carrier correction}
Let $\theta=\omega_0t$, $v=\cos\theta+\sin3\theta$, and $i=g(v)=v+v^3.$
This cubic-resistor example is used to examine reference and projection consistency. $g(v)$ is strictly monotone because $g'(v)=1+3v^2>0$, and it is passive because $vi=v^2+v^4\geq0$. Expanding the cubic term gives current harmonics at $\{1,3,5,7,9\}$, all of which are retained. The corresponding reactive-power components give
\(
Q_1=\frac{3}{8},~Q_3=-\frac{1}{8},~Q_5=Q_7=Q_9=0.
\)
The derivative readout $Q_D=\frac{3}{8}+3\left(-\frac{1}{8}\right)=0$, as required by the periodic memoryless-resistor identity, while the equal-weight readout is $Q_\Sigma=\frac{1}{4}$ for the same data. This result shows that different reactive-power definitions can describe different properties of the same voltage--current pair.

\begin{table}[t]
\caption{E1: integer-frame readouts and carrier recoveries}
\label{tab:E1}\centering
\footnotesize
\begin{tabular}{rrrrr}\toprule
$\rho$ & $Q_{D,\rm rel}^{(\rho)}$ & $\rho\Qsig$ & $Q_D$ & recovered $\rho$\\\midrule
$-3$ & $3/4$ & $-3/4$ & $0$ & $-3$\\
$0$ & $0$ & $0$ & $0$ & $0$\\
$1$ & $-1/4$ & $1/4$ & $0$ & $1$\\
$2$ & $-1/2$ & $1/2$ & $0$ & $2$\\
$3$ & $-3/4$ & $3/4$ & $0$ & $3$\\
$7$ & $-7/4$ & $7/4$ & $0$ & $7$\\
\bottomrule\end{tabular}

\end{table}

Table~\ref{tab:E1} evaluates the same complete data in several integer frames. The columns give the reference shift $\rho$, the relative derivative readout $Q_{D,\rm rel}^{(\rho)}$, the carrier-correction term $\rho Q_\Sigma$, their sum $Q_D$, and the value of $\rho$ recovered from \eqref{eq:resistor}. For example, at $\rho=2$, $Q_{D,\rm rel}^{(\rho)}=-\frac12$ is corrected by $\rho Q_\Sigma=\frac12$ to recover the physical reactive-power readout $Q_D=0$. At $\rho=3$, the third harmonic is shifted to relative zero, $\nu=0$, while its two I/Q coordinates are retained; thus, relative zero does not imply a physical DC component.

Returning to the example in Section I, suppose that an AC circuit with a 60-Hz physical fundamental is measured using a 59.5-Hz processing reference, so that $\rho=\frac{119}{120}$. There are
\(
Q_{D,\mathrm{rel}}^{(119/120)}=-\frac{119}{480}, ~\frac{119}{120}\Qsig=\frac{119}{480}, ~ Q_D=0.
\)
Reporting only the relative derivative without carrier correction would therefore introduce an error of $\frac{-119}{480}$. Adding the correction term recovers the same physical $Q_D=0$, even though the fundamental envelope now varies at $+0.5$ Hz. The equal-weight value remains $\Qsig=\frac14$ because it is a separate readout with constant harmonic weights. 

Since $\Qsig=\frac14\ne0$, this excitation is informative for the conditional recovery formula \eqref{eq:recovery}. Thus, under the assumed memoryless-resistor law, the carrier shift can be recovered from the scalar relation. This example does not establish either the assumed device law or the applicability of the recovery formula to an unknown load.
\vspace{-0.2in}
\subsection{E2: physical weighting at relative zero in an RLC branch}
Consider the periodic series-RLC branch of Section V with $\omega_0=R=L=C=1$ and RMS current phasors $I_1=1$ and $I_3=\frac12$. The voltage phasors are $V_n=[1+\jj(n-\frac{1}{n})]I_n$. Complex powers are $S_n=V_n I_n^*=P_n+\jj Q_n$. Table~\ref{tab:RLC} gives harmonic voltage phasors, harmonic power components and stored energies and related readouts for two frames.

\begin{table}[t]
\caption{E2: harmonic power components and mean stored energies}
\label{tab:RLC}\centering\footnotesize
\begin{tabular}{rrrrrr}\toprule
$n$ & $V_n$ & $P_n$ & $Q_n$ & $W_{L,n}$ & $W_{C,n}$\\\midrule
$1$ & $1$ & $1$ & $0$ & $1/2$ & $1/2$\\
$3$ & $1/2+\jj4/3$ & $1/4$ & $2/3$ & $1/8$ & $1/72$\\
\midrule
$\rho$ & $\nu (n=3)$ & $Q_{D,\rm rel}^{(\rho)}$ & $\rho Q_\Sigma$
& $Q_D$ & $\frac{1}{2(\nu+\rho)}$\\\midrule
$3$ & $0$ & $0$ & $2$ & $2$ & $1/6$\\
$1/2$ & $5/2$ & $5/3$ & $1/3$ & $2$ & $1/6$\\
\bottomrule
\end{tabular}

\end{table}

The total readouts are $P=1+\frac14=\frac54$, $\Qsig=0+\frac23=\frac23$, and $Q_D=0\cdot1+\frac23\cdot3=2$. At $\rho=3$, the only nonzero reactive contribution, $Q_3$, is located at relative order $\nu=n-\rho=3-3=0$, so its relative derivative contribution is $Q_{D,\mathrm{rel}}^{(3)}=0$. Since the carrier contribution is $3\frac23=2$, the physical derivative reactive-power readout remains nonzero.
The physical weight $f(n)=\frac{1}{2\omega_0 n}$ is also finite, $\frac{1}{2\cdot1\cdot(0+3)}=\frac16$, giving the mean storage imbalance $\mathcal E_\Delta=\frac23\cdot\frac16=\frac19$. At the fractional reference $\rho=\frac12$, the relative readout is $\frac53$ and the carrier correction is $\frac13$, again giving $Q_D=2$. Both frames use the same physical order $n=3$ in the energy weight. Finally, the fundamental is at series resonance: $W_{L,1}=W_{C,1}=\frac12$ although $Q_1=0$. Thus, zero signed imbalance does not imply zero total storage.

\vspace{-0.2in}
\subsection{E3: correct weighting with different waveform selections}
Apply a relative-frequency filter having $|\nu|\leq1$ to E1. At integer references $\rho=0,2,3$, it selects physical components $\{1\}$, $\{1,3\}$, and $\{3\}$, respectively. Table~\ref{tab:projection} uses the correct physical weight $\nu+\rho$ in all three calculations. The different reported values are therefore caused by a change in selected physical components, corresponding to $\epsilon_2$ in \eqref{eq:error-decomposition}.

\begin{table}[t]
\caption{E3: projected readouts after correct physical weighting}
\label{tab:projection}\centering
\begin{tabular}{rrrr}\toprule
$\rho$ & physical components & $Q_D$ & normalized current RMS error\\\midrule
$0$ & $\{1\}$ & $3/8$ & $\sqrt{99/188}\approx0.725669$\\
$2$ & $\{1,3\}$ & $0$ & $\sqrt{7/94}\approx\ProjectionRelative$\\
$3$ & $\{3\}$ & $-3/8$ & $\sqrt{103/188}\approx 0.740189$\\
\bottomrule\end{tabular}

\end{table}

Now keep the physical projection $\Pi$ onto $\{1,3\}$ fixed and transport it with each frame. Since $v$ contains only these two harmonics, $\Pi v=v$, while
$ \Pi i =
\frac{13}{4}\cos\theta+\frac34\sin\theta
+\frac14\cos3\theta+\frac{13}{4}\sin3\theta.
$
Thus,
$ i-\Pi i =
-\frac34\cos5\theta+\frac34\sin5\theta
-\frac34\cos7\theta-\frac14\sin9\theta,
$
so the projection omits current harmonics $5,7,9$. If the same physical harmonics $\{1,3\}$ are retained and the reactive-power readout is transformed correctly, changing the reference frame does not change the reported physical value $Q_D=0$. However, the filtered waveforms are not exact. 

\begin{figure}[t]\centering
\includegraphics[width=\columnwidth,height=3.5cm]{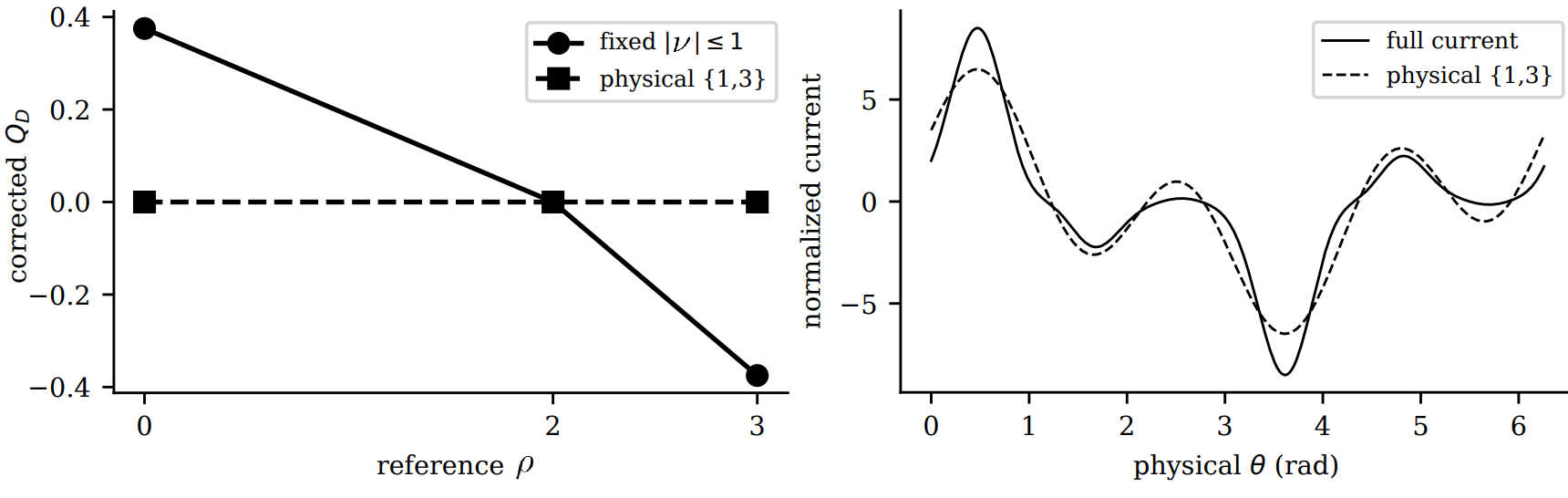}
\caption{E3 projection effects. Left: correctly weighted readouts for a fixed relative filter and a transported physical filter. Right: full current and its projection onto harmonics ${1,3}$; the projection preserves $Q_D=0$ with a normalized RMS current error of $27.3\%$.}
\vspace{-0.2in}
\label{fig:projection}
\end{figure}

For $i_n=A_n\cos n\theta+B_n\sin n\theta$, its squared RMS contribution is $\|i_n\|^2=\frac12(A_n^2+B_n^2)$, giving
\(
\|i_5\|^2=\frac9{16},\qquad
\|i_7\|^2=\frac9{32},\qquad
\|i_9\|^2=\frac1{32}.
\)
Because different harmonics are orthogonal,
\(
\|i-\Pi i\|^2
=
\|i_5\|^2+\|i_7\|^2+\|i_9\|^2
=
\frac78,
\)
while $\|\Pi i\|^2=\frac{87}{8}$ and $\|i\|^2=\frac{47}{4}$. Therefore, the normalized RMS current error is
\(
\frac{\|i-\Pi i\|}{\|i\|}
=
\sqrt{\frac{7}{94}}
\approx\ProjectionRelative.
\)
Since $\Pi v=v$ and the original device law is $g(v)=i$, the projected pair has the same normalized constitutive residual,
$
\frac{\|\Pi i-g(\Pi v)\|}{\|i\|}
=
\sqrt{\frac{7}{94}}
\approx\ProjectionRelative.
$
Thus, the exact scalar agreement $Q_D=0$ coexists with a normalized RMS current error of about $27.3\%$. Figure~\ref{fig:projection} compares the full and projected currents. The omitted current-only harmonics do not contribute to $Q_D$, but they are needed to reconstruct the physical current and satisfy the device law. Corollary~\ref{cor:orbit} shows that the error of this fixed physical projection remains unchanged after any exact carrier shift. The same conclusion holds for general orthogonal projections, not only harmonic filters.
If only harmonic ${1}$ or ${3}$ is selected, $Q_D=\frac38$ or $-\frac38$ and the normalized current error increases to $72.6\%$ or $74.0\%$, respectively, as shown in Table~\ref{tab:projection}.

The analytical examples in this section evaluate deterministic readout consistency rather than instrument performance. Sensor errors, synchronization, quantization, calibration, and stochastic measurement uncertainty are outside the scope of this paper.

\vspace{-0.2in}
\section{Conclusion}
This paper formulates frequency-frame covariance for a specified reactive-power readout. An exact reference change preserves the physical waveform, but preserving the reported scalar also requires transforming the readout operator. For finite harmonic data, frequency-dependent weights must be evaluated at the physical order \(n=\nu+\rho\). Thus, the derivative/Iliovici readout contains both the relative-derivative term and the reference contribution. Omitting the latter produces a deterministic error even with exact waveform coefficients.

The same covariance requirement applies to physical projections. Transforming the projection with the waveform coordinates preserves the selected physical subspace and its approximation error. The nonlinear-resistor example shows that scalar agreement alone does not imply waveform or device-law consistency a projected pair can preserve the derivative reactive-power readout while having nonzero current error and violating the original constitutive relation. The resistor identity also enables conditional recovery of missing carrier information when, with increasing sensitivity to perturbations as  approaches zero. The formulation assumes periodic waveforms, a fixed carrier, and a full-period inner product. Sliding-window, transient, or time-varying-reference formulations require additional endpoint and time-varying-coefficient terms. Future work may examine reference-frequency selection for specific numerical methods while preserving these covariance conditions.

\vspace{-0.2in}
\section*{Acknowledgment}
The author used OpenAI ChatGPT to assist with mathematical derivations and proofs and with editing Sections I--VII. The author reviewed and verified all mathematical developments and conclusions.

\vspace{-0.1in}

\bibliographystyle{IEEEtran}
\bibliography{refs}

@article{Sanders1991,
 author={Sanders, Seth R. and Noworolski, J. Mark and Liu, Xiaojun Z. and Verghese, George C.},
 title={Generalized Averaging Method for Power Conversion Circuits},
 journal={IEEE Trans. Power Electron.}, year={1991}, volume={6}, number={2}, pages={251--259}, month=apr
}

@article{Dinkelbach2021,
 author={Dinkelbach, Jan and Nakti, Ghassen and Mirz, Markus and Monti, Antonello},
 title={Simulation of Low Inertia Power Systems Based on Shifted Frequency Analysis},
 journal={Energies},year={2021},volume={14},number={7},pages={1860}
}

@misc{Jeltsema2014,
 author={Jeltsema, Dimitri and van der Woude, Jacob W. and Hartman, Marek T.},
 title={A Novel Time-Domain Perspective of the {CPC} Power Theory: Single-Phase Systems},
 year={2014},howpublished={arXiv:1403.7842v2}
}

@misc{MontoyaGAPoT2020,
 author={Montoya, Francisco Gil and others},
 title={Geometric Algebra Power Theory ({GAPoT}): Revisiting Apparent Power under Non-Sinusoidal Conditions},
 year={2020},howpublished={arXiv:2002.10011v3}
}

@misc{Sun2026,
 author={Sun, Kai},
 title={Reactive Power Beyond Sinusoids: Scalar Measurement Readouts and Information Limits},
 year={2026},howpublished={arXiv:2609.25329v2}
}

@article{Orallo2014,
 author={Orallo, Carlos M. and Carugati, Ignacio and Maestri, Sebastian and Donato, Patricio G. and Carrica, Daniel and Benedetti, Mario},
 title={Harmonics Measurement With a Modulated Sliding Discrete Fourier Transform Algorithm},
 journal={IEEE Trans. Instrum. Meas.},
 year={2014}, volume={63}, number={4}, pages={781--793}
}

@article{GarciaVeloso2024,
 author={Garc{\'i}a-Veloso, C{\'e}sar and Paolone, Mario and Maza-Ortega, Jos{\'e} Mar{\'i}a},
 title={Computationally Efficient Synchrophasor Estimation: Delayed In-Quadrature Interpolated {DFT}},
 journal={IEEE Trans. Instrum. Meas.},
 year={2024}, volume={73}, pages={1--16}
}

@article{Macii2025,
 author={Macii, David and Pegoraro, Paolo Attilio and Sitzia, Carlo},
 title={Clustering and Fusion of Distributed Multitone Frequency Measurement Data for Enhanced Grid-Level Power Quality Monitoring},
 journal={IEEE Trans. Instrum. Meas.},
 year={2025}, volume={74}, pages={1--14}
}

@misc{IEEE1459_2025,
 author={{IEEE}},
 title={{IEEE} Standard Definitions for the Measurement of Electric Power Quantities Under Sinusoidal, Nonsinusoidal, Balanced, or Unbalanced Conditions},
 howpublished={IEEE Std 1459-2025}, year={2025}
}

@article{Stankovic2000,
  author={Stankovic, Aleksandar M. and Aydin, Timur},
  title={Analysis of Asymmetrical Faults in Power Systems Using Dynamic Phasors},
  journal={IEEE Trans. Power Syst.},
  year={2000},
  volume={15},
  number={3},
  pages={1062--1068}
}

@article{Stankovic2002,
  author={Stankovi{\'c}, Aleksandar M. and Sanders, Seth R. and Aydin, Timur},
  title={Dynamic Phasors in Modeling and Analysis of Unbalanced Polyphase AC Machines},
  journal={IEEE Trans. Energy Conv.},
  year={2002},
  volume={17},
  number={1},
  pages={107--113},
  month=mar
}

@article{Iliovici1925,
  author  = {Iliovici, M.},
  title   = {D{\'e}finition et mesure de la puissance et de l'{\'e}nergie r{\'e}actives},
  journal = {Bulletin de la Soci{\'e}t{\'e} Fran{\c c}aise des {\'E}lectriciens},
  volume  = {5},
  pages   = {931--956},
  year    = {1925}
}

@inproceedings{Hartman2009,
  author    = {Marek T. Hartman},
  title     = {The Integral Method to Calculate the Power States in Electrical Circuits},
  booktitle = {Proc. 6th Int. Conf. Compatibility and Power Electronics (CPE)},
  pages     = {174--179},
  year      = {2009},
  doi       = {10.1109/CPE.2009.5156032}
}
\end{document}